\documentclass[%
 aip,
 amsmath,amssymb,
 reprint,
 twocolumn,
 dblfloatfix]{revtex4-2}
\usepackage{graphicx}% Include figure files
\usepackage{dcolumn}% Align table columns on decimal point
\usepackage{bm}% bold math
\usepackage{siunitx}
\usepackage{amsmath}

\usepackage[usenames,dvipsnames,svgnames,table]{xcolor}
\colorlet{m}{black}
\colorlet{k}{Black}
\usepackage[textwidth=12.5mm]{todonotes}
\usepackage{easyReview}
\usepackage[colorlinks=true,citecolor=blue]{hyperref}
\usepackage[rightcaption]{sidecap}

\begin{document}

\preprint{AIP/123-QED}

\title{Shubnikov-de Haas oscillations in coherently strained AlN/GaN/AlN quantum wells on bulk AlN substrates
}% Force line breaks with \\
%\thanks{Footnote to title of article.}

% \title{Coherently strained high conductivity AlN/GaN/AlN quantum-well HEMT heterostructures on single-crystal AlN substrates with silicon delta doping
% }% Force line breaks with \\
% %\thanks{Footnote to title of article.}

\author{Yu-Hsin Chen}
\email{yc794@cornell.edu}
\affiliation{\hbox{Department of Materials Science and Engineering, Cornell University, Ithaca, NY, 14853, USA}}

\author{Jimy Encomendero}%
\affiliation{\hbox{School of Electrical and Computer Engineering, Cornell University, Ithaca, NY, 14853, USA}}

\author{Huili Grace Xing}
\affiliation{\hbox{Department of Materials Science and Engineering, Cornell University, Ithaca, NY, 14853, USA}}
\affiliation{\hbox{School of Electrical and Computer Engineering, Cornell University, Ithaca, NY, 14853, USA}}
\affiliation{\hbox{Kavli Institute at Cornell for Nanoscale Science, Cornell University, Ithaca, NY, 14853, USA}}

\author{Debdeep Jena}
\email{djena@cornell.edu}
\affiliation{\hbox{Department of Materials Science and Engineering, Cornell University, Ithaca, NY, 14853, USA}}
\affiliation{\hbox{School of Electrical and Computer Engineering, Cornell University, Ithaca, NY, 14853, USA}}
\affiliation{\hbox{Kavli Institute at Cornell for Nanoscale Science, Cornell University, Ithaca, NY, 14853, USA}}

\begin{abstract}
We report the observation of Shubnikov-de Haas (SdH) oscillations in coherently strained, low-dislocation AlN/GaN/AlN quantum wells (QWs), including both undoped and $\delta$-doped structures. SdH measurements reveal a single subband occupation in the undoped GaN QW and two subband occupation in the $\delta$-doped GaN QW. More importantly, SdH oscillations enable direct measurement of critical two-dimensional electron gas (2DEG) parameters at the Fermi level: carrier density and ground state energy level, electron effective mass ($m^* \approx 0.289\,m_{\rm e}$ for undoped GaN QW and $m^* \approx 0.298\,m_{\rm e}$ for $\delta$-doped GaN QW), and quantum scattering time ($\tau_{\rm q} \approx 83.4 \, \text{fs}$ for undoped GaN QW and $\tau_{\rm q} \approx 130.6 \, \text{fs}$ for $\delta$-doped GaN QW). These findings provide important insights into the fundamental properties of 2DEGs that are strongly quantum confined in the thin GaN QWs, essential for designing nitride heterostructures for high-performance electronic applications.

\end{abstract}

\maketitle

The discovery of polarization-induced two-dimensional electron gases (2DEGs) in AlGaN/GaN heterojunction in the 1990s revolutionized the development of nitride-based high-electron mobility transistors (HEMTs).\cite{khan1992observation,asif1993high} These GaN-based devices, characterized by their wide bandgaps, high electron velocities, and ability to form heterojunctions with high carrier concentration and mobility, have since become essential for high-frequency RF amplifiers\cite{mishra2008gan,mishra2002algan} and fast high-voltage switching applications\cite{amano20182018,flack2016gan}.

More recently, 2DEGs have been demonstrated in thin, coherently strained GaN quantum wells (QWs) sandwiched between AlN buffers and AlN top barriers.\cite{qi2017strained,chaudhuri2022integrated,chen2024electron,chen2024high} These AlN/GaN/AlN heterostructures, epitaxially grown on single-crystal AlN substrates, achieve a five-order reduction in dislocation density (DD) down to \(\approx 10^4 \, \text{cm}^{-2}\).\cite{bondokov2021two} While the AlN layers are strain-free due to lattice-matching with the AlN substrate, the thin GaN channel is under a large compressive strain of \(-2.4\%\), which alters its electronic band structure. The large discontinuity in spontaneous and piezoelectric polarization between GaN and AlN induces a high-density 2DEG, and the large energy band offset confines the 2DEG within the QW. Such GaN QW heterostructures provide a 2-3 times higher 2DEG densities (\(n_{\rm s} \sim 2-3 \times 10^{13} \, \text{cm}^{-2}\)) compared to conventional AlGaN/GaN heterostructures (\(n_{\rm s} \sim 10^{12}-10^{13} \, \text{cm}^{-2}\)), enabling access to a larger \(k\)-space in the GaN conduction band.

A key characteristic of high-quality 2DEGs is the occurrence of quantum oscillations in longitudinal magnetoresistance \(R_{\rm xx}\), known as Shubnikov–de Haas (SdH) oscillations. For these quantization effects to manifest in a magnetic field, the cyclotron energy \(\hbar \omega_{\rm c}\) must exceed both the thermal energy \(k_{\rm b}T\) and the Landau level broadening \(\hbar / \tau_{\rm q}\)\cite{sladek1958magnetoresistance}, where \(\omega_{\rm c} = eB \, / \, m^*\) is the cyclotron frequency, \(k_{\rm b}\) is the Boltzmann constant, \(\tau_{\rm q}\) is the quantum scattering time, and \(m^*\) is the electron effective mass. This condition indicates that observing quantum oscillations necessitates the application of a strong magnetic field, and the use of high-quality samples with minimal scattering. In our recent studies, we improved the growth process for AlN/GaN/AlN heterostructures and enhanced the channel mobility through silicon $\delta$-doping.\cite{chen2024electron,chen2024high} Building on this progress, here we report a magnetotransport study of both the undoped and $\delta$-doped GaN QWs, revealing well-resolved SdH oscillations in \(R_{\rm xx}\) of both structures. These quantum oscillations allow us to extract critical parameters for 2D carrier transport, including (1) the carrier density and ground state energy levels, (2) the electron effective mass, and (3) the quantum scattering time.

%----------------------------------------------------
\begin{figure*}
	\centering
	\includegraphics[width=1\textwidth]{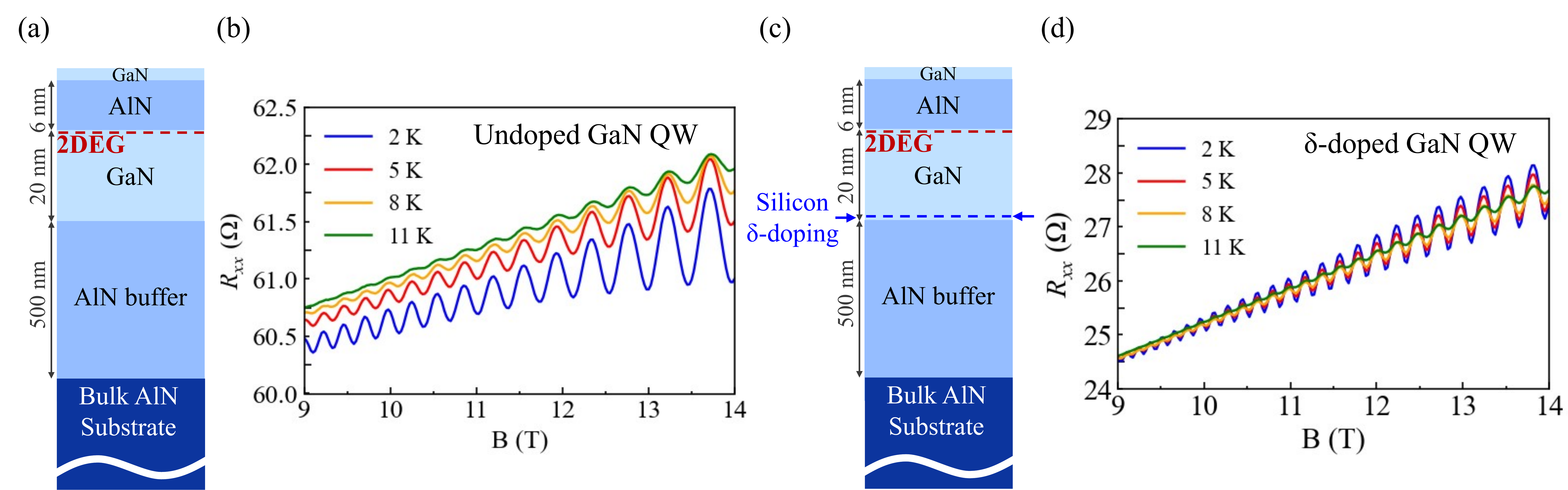} % Here is how to import EPS art
	\caption{\label{fig:1} (a) Schematic of the epitaxial undoped AlN/GaN/AlN QW heterostructure. 
	(b) Measured longitudinal magnetoresistance \(R_{\rm xx}\) of 2DEG in the undoped GaN QW as a function of magnetic field \(B\) from 9 to 14 T, at various temperatures ranging from 2 to 11 K. 
	(c) Schematic of the $\delta$-doped AlN/GaN/AlN QW, which includes a sheet of silicon donors with a doping density of \(\sigma_{\delta} = 5 \times 10^{13}\,\text{cm}^{-2}\) at the bottom GaN/AlN interface, approximately 26 nm below the surface. 
	(d) Measured \(R_{\rm xx}\) of 2DEG in the $\delta$-doped GaN QW within the same magnetic field and temperature range, showing more frequent oscillations.
	}
\end{figure*}
%----------------------------------------------------
\begin{figure*}
	\centering
	\includegraphics[width=0.992\textwidth]{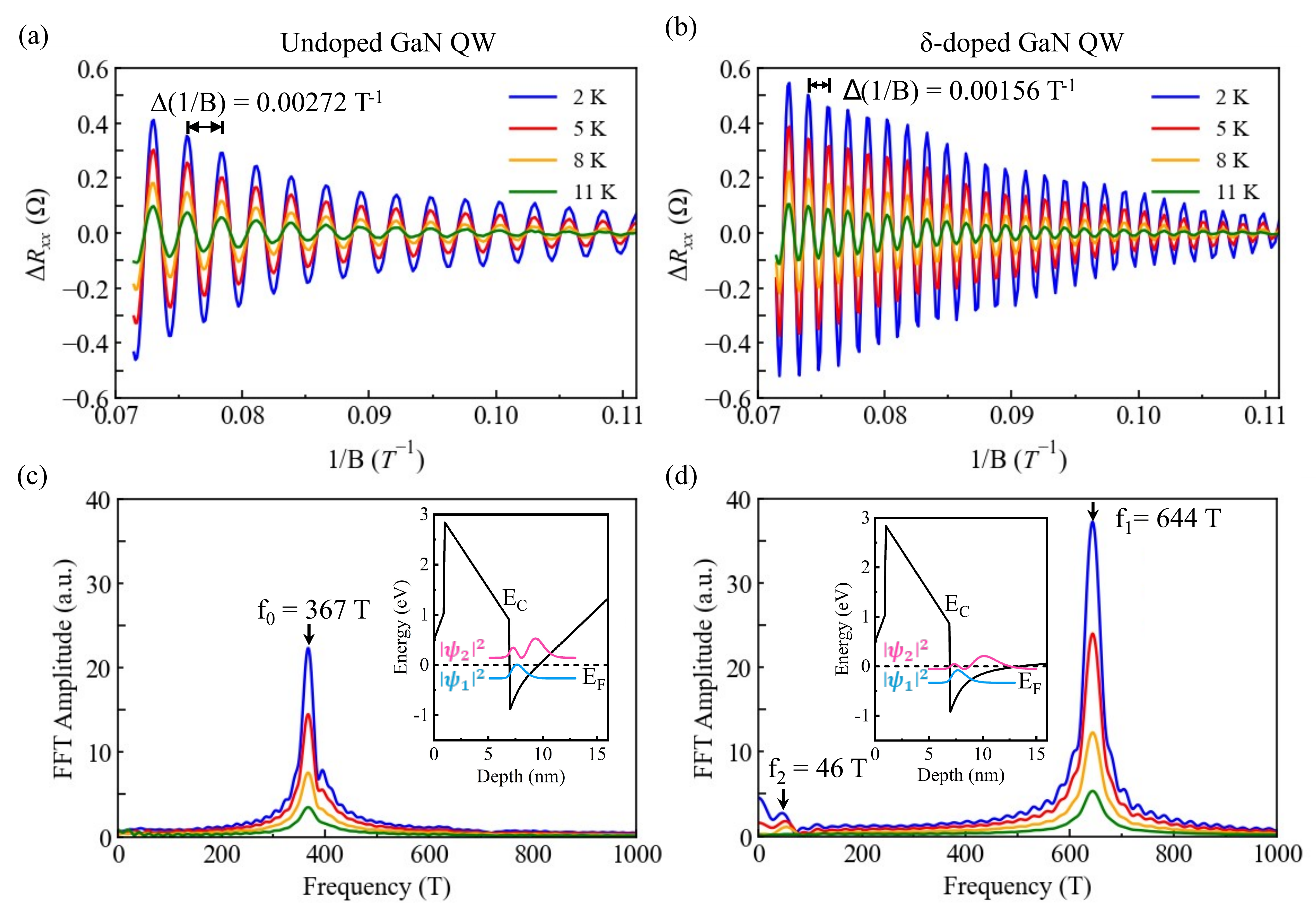} 
	\caption{\label{fig:2} (a) The oscillatory component \(\Delta R_{\rm xx}\) of the undoped GaN QW plotted against \(1/B\). The oscillations are periodic in \(\Delta(1/B) = 0.00272 \, \text{T}^{-1}\), corresponding to a 2DEG density of \(n_{\rm s} = 1.78 \times 10^{13} \, \text{cm}^{-2}\).
	(b) \(\Delta R_{\rm xx}\) of the $\delta$-doped GaN QW is periodic in \(\Delta(1/B) = 0.00156 \, \text{T}^{-1}\), corresponding to a 2DEG density of \(n_{\rm s} = 3.11 \times 10^{13} \, \text{cm}^{-2}\).
	(c) The fast Fourier transform (FFT) spectrum of the oscillating \(\Delta R_{\rm xx}\) versus 1/B in the undoped GaN QW shows a single frequency at $f_0$ = 367 T, indicating single subband occupation.
	(d) The FFT spectrum for the $\delta$-doped GaN QW reveals two distinct frequency peaks at $f_1$ = 644 T and $f_2$ = 46 T, indicating two subband occupation. The insets in (c) and (d) are the calculated conduction band energy profiles and the squared-amplitude of the electronic wavefunctions for both structures, respectively.}
\end{figure*}
%----------------------------------------------------
\begin{figure}[ht]
	\centering
	\includegraphics[width=1\columnwidth]{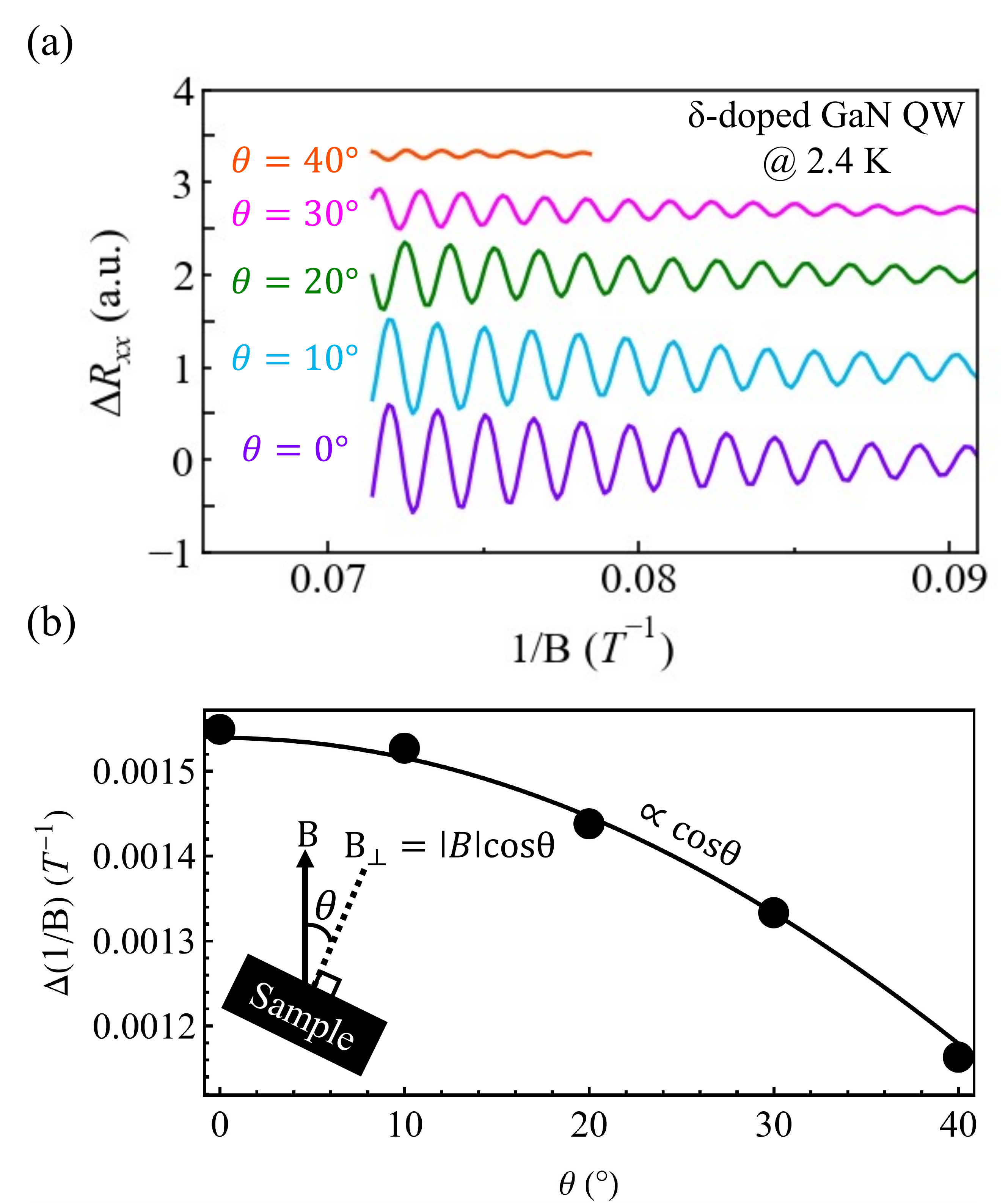}
	\caption{(a) SdH oscillations measured at various tilt angles for the $\delta$-doped GaN QW at 2.4 K. 
    The angle \(\theta\) represents the tilt of the sample surface normal relative to the applied magnetic field direction, as shown in the inset of Fig. 2(b).
	(b) The dependence of the oscillation period \(\Delta(1/B)\) on \( \cos \theta \) confirms the two-dimensional confinement of electrons in the $\delta$-doped GaN QW.}
	\label{fig:3}
\end{figure}
%----------------------------------------------------
\begin{figure*}
	\centering
	\includegraphics[width=0.7\textwidth]{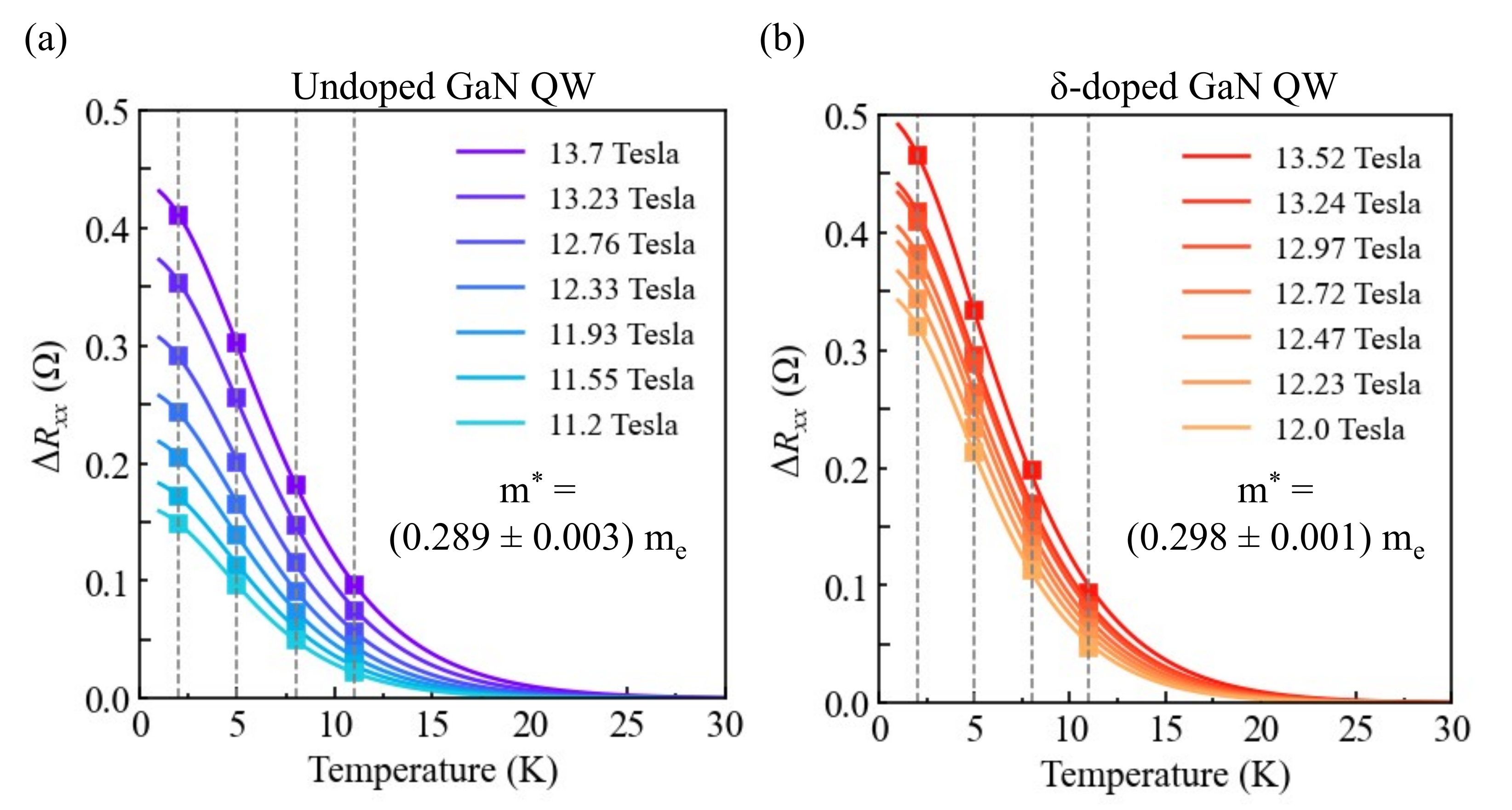}
	\caption{\label{fig:4} Temperature dependence of \(\Delta R_{\rm xx}\) at various B fields for (a) undoped GaN QW and (b) the first subband of \(\delta\)-doped GaN QW. Each solid curve represents the fits to \(\chi / \sinh(\chi)\) in Eq.(1) at a certain B field to extract the electron effective mass. Curve fittings of the thermal damping \(\Delta R_{\rm xx}\) yield an electron effective mass of \( m^* = (0.289 \pm 0.003) \, m_{\rm e} \) for the undoped GaN QW and \( m^* = (0.298 \pm 0.001) \, m_{\rm e} \) for the first subband of \(\delta\)-doped GaN QW. The reported electron effective mass and its error represent the mean and standard deviation of the numerically fitted values obtained across different magnetic fields.}
\end{figure*}

%----------------------------------------------------

Figures~\ref{fig:1}(a) and ~\ref{fig:1}(c) present a schematic of the undoped and $\delta$-doped GaN QW heterostructures used in this study. Both samples were grown epitaxially on single-crystal AlN substrates using molecular beam epitaxy. The detailed growth procedure, as well as the structural and transport characteristics of these two samples, have been thoroughly described in our recent work.\cite{chen2024high} Low-field Hall-effect measurements show that the undoped GaN QW exhibits a 2DEG density of \(2.04 \times 10^{13} \, \text{cm}^{-2}\) with electron mobility of \(611.1 \, \text{cm}^2/\text{Vs}\) at room temperature (RT) and \(1210.4 \, \text{cm}^2/\text{Vs}\) at 2 K. On the other hand, the $\delta$-doped GaN QW shows a higher 2DEG density of \(3.24 \times 10^{13} \, \text{cm}^{-2}\) with enhanced mobilities of \(854.6 \, \text{cm}^2/\text{Vs}\) at RT and \(2234.6 \, \text{cm}^2/\text{Vs}\) at 2 K. Both samples exhibit a temperature-independent 2DEG density down to cryogenic temperatures, confirming their polarization-induced nature.

Magnetotransport measurements were performed on both heterostructures using ohmic contacts formed in a van der Pauw configuration. The samples were placed in a Physical Property Measurement System (PPMS) from Quantum Design with a DC excitation current of \(100 \, \mu\text{A}\). Figures.~\ref{fig:1}(b) and ~\ref{fig:1}(d) show the \(R_{\rm xx}\) as a function of the magnetic field \(B\) applied perpendicular to the sample surface, across a range of 9 to 14 T and measured at various temperatures from 2 to 11 K. As the magnetic field increases, the amplitude of the oscillations grows due to the larger separation between Landau levels and the increased number of states within each level, causing greater fluctuations in the density of states. In contrast, the weakening of oscillation amplitude with increasing temperature is attributed to the smearing of the Fermi–Dirac distribution near the Fermi level.\cite{zhang2023tunable} The onsets of the SdH oscillations were recorded at around 8 T for both samples, as shown in Fig. S1 in the supplementary material, significantly lower than similar QW heterostructures grown on SiC substrates, where over 25 T is required to resolve quantum oscillations.\cite{chaudhuri2022integrated}  

To isolate the oscillatory component \(\Delta R_{\rm xx}\), the background resistance was subtracted from the measured \(R_{\rm xx}\). This background is determined by averaging polynomial fits applied separately to the peaks and valleys of the SdH oscillations\cite{okazaki2018shubnikov}, as shown in Fig. S2 of the supplementary material file. Figures~\ref{fig:2}(a) and ~\ref{fig:2}(b) show the resulting \(\Delta R_{\rm xx}\) as a function of \(1/B\) for both samples, highlighting the periodic nature of the SdH oscillations in \(1/B\). The oscillatory component \(\Delta R_{\rm xx}\) is given by\cite{jena2022quantum,ihn2009semiconductor}:
\begin{equation}
\Delta R_{\rm xx} \propto \frac{\chi}{\sinh(\chi)}e^{-\frac{\pi}{\omega_{\rm c} \tau_{\rm q}}}\cos\left(\frac{2\pi E_{\rm F}}{\hbar \omega_{\rm c}}\right),
\end{equation} where \(\chi = 2\pi^2 k_B T / (\hbar \omega_{\rm c})\) and \(E_{\rm F} = n_{\rm s} \pi \hbar^2 / m^*\) is the Fermi level of 2DEG for a particular subband. The period of the oscillation \(\Delta(1/B)\) is determined by the cosine term in Eq. (1) and directly measures the 2DEG density \(n_{\rm s}\) via the relation \(n_{\rm s} = q / (\Delta(1/B)\pi \hbar)\). For the undoped GaN QW shown in Fig.~\ref{fig:2}(a), the oscillation periodicity is extracted as \(\Delta(1/B) = 0.00272 \, \text{T}^{-1}\), corresponding to a 2DEG density of \(n_{\rm s} = 1.78 \times 10^{13} \, \text{cm}^{-2}\). For the $\delta$-doped GaN QW presented in Fig.~\ref{fig:2}(b), one can observe that a higher frequency oscillation is superimposed on a lower frequency oscillation. The high-frequency oscillation, with \(\Delta(1/B) = 0.00156 \, \text{T}^{-1}\), is associated with electrons occupying the first subband with \(n_{\rm s} = 3.11 \times 10^{13} \, \text{cm}^{-2}\). The low-frequency oscillation is attributed to electrons in the second subband, which requires further analysis using fast Fourier transform (FFT) to extract its 2DEG density.  The 2DEG densities extracted from \(\Delta(1/B)\) exhibit an error margin of 5\%.

Figures~\ref{fig:2}(c) and ~\ref{fig:2}(d) present the FFT analysis of $\Delta R_{\rm xx}$ versus \(1/B\) for both samples, considering the field interval from 9 to 14 T. The insets in Fig.(2) show the calculated conduction band energy profiles and the squared-amplitude of the electronic wavefunctions for both structures, obtained using a self-consistent Schrödinger–Poisson solver. These calculations indicate that the undoped GaN QW hosts electrons in a single subband, whereas the $\delta$-doped GaN QW has two populated subbands.

From the FFT spectrum in Fig.~\ref{fig:2}(c), a single oscillation frequency at $f_0$ = 367 T is identified for the undoped GaN QW, confirming single subband occupation. The FFT frequency directly measures the 2DEG density, which is given by the Onsager relation: \(n_{\rm s} = \left( 2q / h \right) \times f\), yielding $n_{\rm s} = 1.77 \times 10^{13} \, \text{cm}^{-2}$. This result agrees well with the electronic density obtained both from the real space analysis and low-field Hall effect measurement. In contrast to the undoped structure, the $\delta$-doped GaN QW shows two distinct frequency peaks in the FFT spectrum, indicating the presence of two occupied subbands [See Fig. 2(d)]. The peak at $f_1$ = 644 T corresponds to the first subband with $n_{\rm s1} = 3.11 \times 10^{13} \, \text{cm}^{-2}$, matching the density extracted from $\Delta(1/B)$. The second peak at $f_2$ = 46 T corresponds to electrons in the second subband with $n_{\rm s2} = 2.22 \times 10^{12} \, \text{cm}^{-2}$. The total 2DEG density $n_{\rm s} = n_{\rm s1} + n_{\rm s2} = 3.33 \times 10^{13} \, \text{cm}^{-2}$ agrees well with both the low-field Hall-effect measurement and self-consistent Schrödinger–Poisson calculations.

The Landau quantization of a 2D electron system is only due to the component of the magnetic field perpendicular to the 2D plane, \( B_\perp\). By rotating the angle \(\theta\) between the B field vector and the 2D plane, the period of oscillation changes according to \( B_\perp = |B| \cos\theta \). Fig.~\ref{fig:3}(a) presents the angular dependence of the SdH oscillations measured in the $\delta$-doped GaN QW at 2.4 K. This measurement involved rotating the normal of the 2D plane by an angle \(\theta\) away from the direction of the applied B field, as depicted in the inset of Fig.~\ref{fig:3}(b). Figure~\ref{fig:3}(b) shows that the oscillation period \(\Delta(1/B)\) exhibits a cosine dependence on the rotation angle \(\theta\), indicating that the oscillations solely arise from the perpendicular component of the applied magnetic field $B_\perp$. This characteristic behavior confirms, experimentally, the two-dimensional nature of the electron gas in the $\delta$-doped GaN QW heterostructure.

By analyzing the temperature-dependent SdH oscillation at a fixed B field, the effective mass of the 2DEG at the Fermi level can be determined. To achieve this, \( R_{\rm xx} \) was measured at four different temperatures ranging from 2 to 11 K. Figure~\ref{fig:4}(a) shows the temperature dependent \(\Delta R_{\rm xx}\) for the undoped GaN QW at various B fields. For each B field, the measured \(\Delta R_{\rm xx}\) peak values are overlaid with the best fit to the thermal damping term \(\chi/\sinh(\chi)\) from Eq.(1). Numerical fitting yields an electron effective mass of \( m^* = (0.289 \pm 0.003) \, m_{\rm e} \) for the undoped GaN QW, where \( m_{\rm e} \) is the free electron mass. The reported  electron effective mass and its error represent the mean and standard deviation of the numerically fitted values obtained across different magnetic fields. A similar analysis was done for the $\delta$-doped GaN QW. However, owing to its double subband occupation, a clear beating pattern is discernible in \( \Delta R_{\rm xx} \) [See Fig.~\ref{fig:2}(b)]. This beating arises from the superposition of two oscillatory components corresponding to each electronic subband. To deconvolve each oscillatory component, an inverse FFT algorithm was applied to the individual peak of the FFT spectrum. This procedure is illustrated in detail in Fig. S3 of the supplementary material. After isolating \( \Delta R_{\rm xx} \) for the first subband, its temperature dependence was analyzed to extract the electron effective mass. Numerical fitting yields \( m^* = (0.298 \pm 0.001) \, m_{\rm e} \) for the first subband of the $\delta$-doped GaN QW, as shown in Fig.~\ref{fig:4}(b). In contrast, the oscillatory component corresponding to the second subband is too weak for reliable mass extraction. To improve the precision of electron effective mass extraction in GaN QWs, future work will prioritize measurements at higher magnetic fields and across a broader temperature range.

The extracted effective masses of the GaN QWs in this study are slightly higher than the typical value of \( m^* \approx 0.2 - 0.23 \, m_{\rm e} \) measured in conventional AlGaN/GaN heterostructures\cite{wong1998magnetotransport,elhamri19980,saxler2000characterization,wang2000magnetotransport}, where the GaN layer is relaxed and the 2DEG densities are lower, around \( n_{\rm s} \sim 5 \times 10^{12} \, \text{cm}^{-2} \). Several factors could contribute to these higher effective mass, including : (1) the compressive strain experienced by the GaN QW on AlN, which increases the in-plane effective mass\cite{dreyer2013effects}; (2) the high 2DEG densities, which push the Fermi energy further away from the subband minimum, thereby amplifying non-parabolicity effect.\cite{syed2003nonparabolicity}; (3) quantum confinement, which could also play a role in enhancing the effective mass.\cite{kurakin2009quantum} A detailed theoretical study to fully understand the effective mass in GaN QWs remains an important area for future work.

%----------------------------------------------------
\begin{figure}[ht]
	\centering
	\includegraphics[width=0.8\columnwidth]{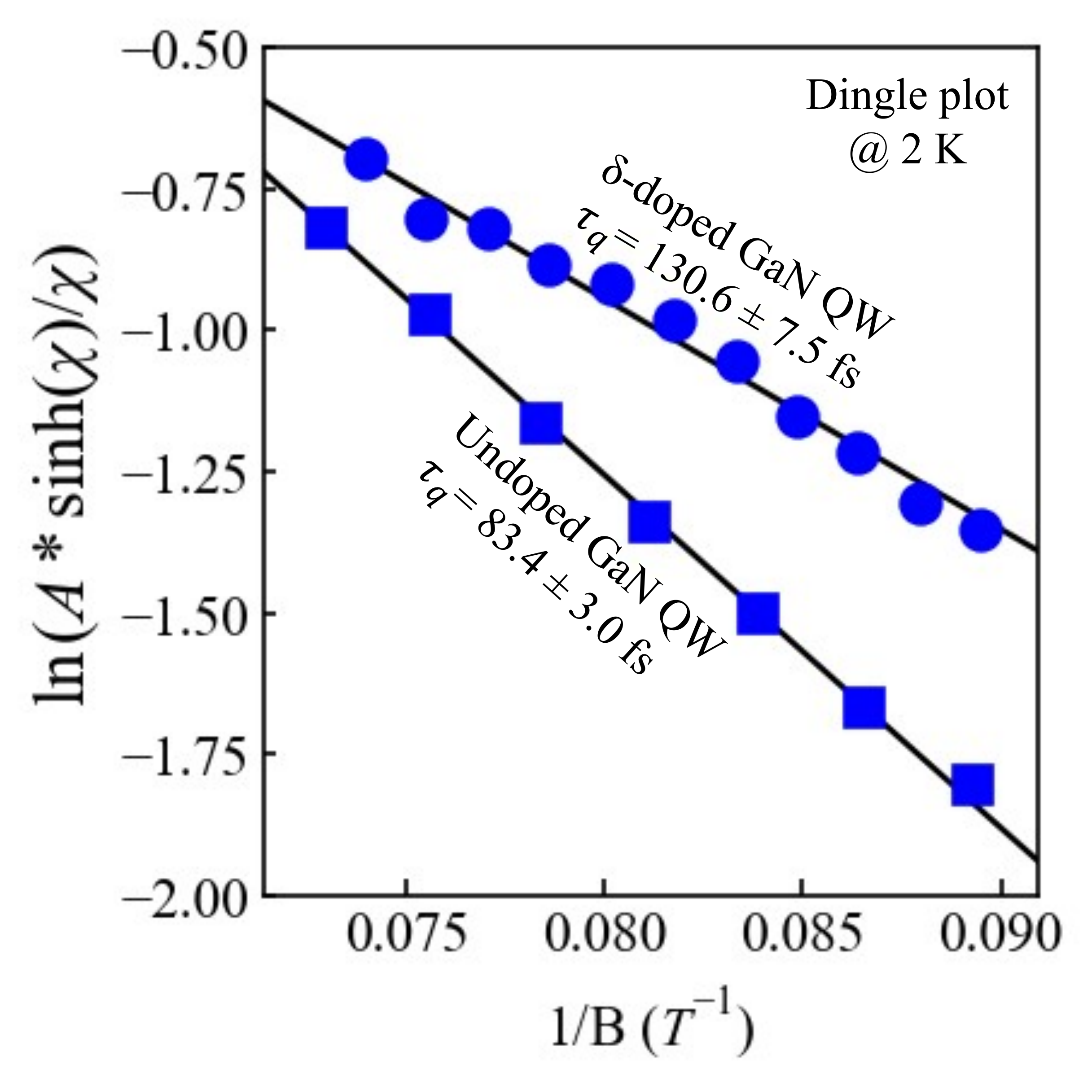}
	\caption{ The quantum scattering time \(\tau_{\rm q}\) is extracted from the damping of the oscillation amplitudes as a function of \(1/B\). Shown here is the Dingle plot, where \(\ln\left(A^* \sinh(\chi) \, / \, \chi\right)\) is plotted against \(1/B\) at 2 K for the first subband of $\delta$-doped (blue circle) and undoped (blue square) GaN QWs. \( A^* \) is the peak values of \( \Delta R_{\rm xx} \). The solid black lines are fits to the disorder damping term \(e^{-\pi / (\omega_{\rm c} \tau_{\rm q})}\) from Eq.(1), yielding $\tau_{\rm q} = 130.6\pm7.5\,\text{fs}$ for the first subband of $\delta$-doped GaN QW and $\tau_{\rm q} = 83.4\pm3.0\,\text{fs}$ for the undoped GaN QW.}
	\label{fig:5}
\end{figure}
%----------------------------------------------------

The experimental measurement of both the effective mass \( m^* \) and the 2DEG density \( n_{si} \) for each subband allow us to accurately determine the electronic spectrum in each heterostructure. To do so, the subband population can be expressed as \(n_{si} = g_{2D}(E_F - E_i)\), where \( E_F \) is the Fermi energy, \( E_i \) is the subband energy, and \(g_{2D}\) is the two-dimensional density of states, given by \( g_{2D} = m^*/\pi\hbar^2 \). From this analysis, we obtain that \( E_F - E_1 \approx 145 \, \text{meV} \) for the undoped GaN QW, and \( E_F - E_1 \approx 248 \, \text{meV} \) for the $\delta$-doped GaN QW. The lower subband energy in the $\delta$-doped GaN QW results from a reduced internal electric field in the well, achieved by incorporating $\delta$-doping, thus compensating the 2DHG at the bottom GaN/AlN interface. Additionally, the intersubband transition energy in the $\delta$-doped GaN QW is \(E_2 - E_1 \approx 230 \, \text{meV},\) indicating that the second subband is relatively shallow (only 18 meV below \(E_F)\), with most electrons residing in the first subband.

Owing to the presence of disorder, Landau levels exhibit a finite broadening due to their coupling with various scattering potentials.  This broadening can be experimentally quantified by the quantum scattering lifetime \( \tau_{\rm q} \), which represents the mean time that electrons remain in one quantized orbit before scattering into another. Experimentally, \( \tau_{\rm q} \) determines the exponential increase of the amplitude of \( \Delta R_{\rm xx} \), as the reciprocal of the magnetic field (1/B) decreases [See Figs. 2(a) and (b)]. At a given temperature, \( \tau_{\rm q} \) is determined from the Dingle plot, where \(\ln\left(A^* \sinh(\chi) \, / \, \chi\right)\) is plotted against \(1/B\) and \( A^* \) represents the peak values of \(\Delta R_{\rm xx}\) [See Fig.~\ref{fig:5}]. \( \tau_{\rm q} \) is then obtained by fitting the data points versus \(1/B\), using the disorder damping term \(e^{-\pi / (\omega_{\rm c} \tau_{\rm q})}\) from Eq.(1). As shown in Fig.~\ref{fig:5}, \( \tau_{\rm q} \) of the undoped GaN QW is determined to be $\tau_{\rm q} = 83.4\pm3.0\,\text{fs}$ at 2 K. For the $\delta$-doped GaN QW, precise determination of \( A^* \) was attained by isolating \( \Delta R_{\rm xx} \) of the first subband, as previously described and shown in Fig. S3. Using this approach, the quantum lifetime $\tau_{\rm q} = 130.6\pm7.5\,\text{fs}$ was extracted for the first subband of the $\delta$-doped GaN QW.

The quantum scattering lifetime \( \tau_{\rm q} \) accounts for all the scattering mechanisms equally, in contrast, the momentum scattering time \( \tau_{\rm m} \) favors large-angle scattering over small angle scattering due to the weighting factor of \( (1 - \cos \theta) \), where \( \theta \) is the scattering angle.\cite{harrang1985quantum} Experimentally, \( \tau_{\rm m} \) is determined from the low-field Hall mobility via the Drude relation \(\mu = e \tau_{\rm m} / m^*.\) At 2 K, Hall-effect mobility measurements yield \( \tau_{\rm m}\) = 198.9 fs for the undoped GaN QW and \( \tau_{\rm m}\) = 378.6 fs for the $\delta$-doped GaN QW. The Dingle ratios \( \tau_{\rm m}/\tau_{\rm q} \sim 2-3 \), being close to unity for both samples, suggest the prevalence of short-range isotropic scattering potentials, likely due to temperature-independent interface roughness (IR) scattering\cite{harrang1985quantum,hsu2002transport}, which was previously identified as the primary scattering mechanism at cryogenic temperatures for both AlN/GaN/AlN QW heterostructures.\cite{chen2024high}

The electron effective mass and quantum lifetime can also be extracted by directly calculating the SdH oscillations and identifying the combination of \( m^*\) and \( \tau_{\rm q} \) that minimizes the root mean square (RMS) error relative to the measured \( \Delta R_{\rm xx} \) oscillations. This method was applied to the undoped GaN QW, yielding the lowest RMS when \( m^* = 0.292\, m_{\rm e} \) and $\tau_{\rm q}$ = 84~fs, which closely matches the values extracted from fitting the thermal damping \( \Delta R_{\rm xx} \) and the Dingle plot.

In summary, 2DEGs in coherently strained undoped and $\delta$-doped AlN/GaN/AlN heterostructures, grown on low dislocation density single-crystal AlN substrates, exhibit SdH oscillations. FFT analysis of SdH oscillations confirms a single subband occupation in the undoped GaN QW, and double subband occupation in the $\delta$-doped GaN QW. Analysis of their thermal damping allows us to directly measure electron effective mass, yielding \( m^* \approx 0.289 \, m_{\rm e} \) for the undoped GaN QW and \( m^* \approx 0.298 \, m_{\rm e} \) for the $\delta$-doped GaN QW, highlighting a fundamental distinction compared to AlGaN/GaN heterostructures with \( m^* \approx 0.2 - 0.23 \, m_{\rm e} \). Furthermore, analysis of the Dingle plot reveals a longer quantum scattering time in the $\delta$-doped GaN QW, attributed to the lowering of interface roughness scattering, enabled by $\delta$-doping. The experimental measurement of these fundamental transport properties is essential not only for designing heterostructures with improved electronic transport, but also for enhancing the accuracy of device modeling. These insights hold significant technological and scientific value for advancing nitride-based electronic devices.

\section*{Supplementary Material}
Refer to the supplementary material for (1) \(R_{\rm xx}\) measurements across the full magnetic field range, including a focus on the SdH oscillation onsets, (2) details on background resistance subtraction to obtain \(\Delta R_{\rm xx}\), and (3) The procedure to deconvolve two oscillatory components in \(\Delta R_{\rm xx}\) of $\delta$-doped GaN QW.

\begin{acknowledgments}
The authors would like to thank Xiaoxi Huang, Chuan Chang, and Anand Ithepalli for their assistance with wire bonding at Cornell University. This work is supported by Army Research Office under Grant No. W911NF2220177 (Characterization); and ULTRA, an Energy Frontier Research Center funded by the U.S. Department of Energy (DOE), Office of Science, Basic Energy Sciences (BES), under Award No. DE-SC0021230 (Modeling); and DARPA THREADS program (Epitaxial growth). This work made use of the Cornell Center for Materials Research shared instrumentation facility which are supported through the NSF MRSEC program (DMR-1719875), and Kavli Institute at Cornell (KIC). 

\end{acknowledgments}

\section*{Author Declarations}
\subsection*{Conflict of Interest}
The authors have no conflicts to disclose.

\section*{Data Availability}
The data that support the findings of this study are available from the corresponding author upon reasonable request.

\section*{References}
\bibliography{main}% Produces the bibliography via BibTeX.

\end{document}

\preprint{AIP/123-QED}

\title{Shubnikov-de Haas oscillations in coherently strained AlN/GaN/AlN quantum wells on bulk AlN substrates}% Force line breaks with \\
%\thanks{Footnote to title of article.}

\author{Yu-Hsin Chen}
\email{yc794@cornell.edu}
\affiliation{\hbox{Department of Materials Science and Engineering, Cornell University, Ithaca, NY, 14853, USA}}

\author{Jimy Encomendero}%
\affiliation{\hbox{School of Electrical and Computer Engineering, Cornell University, Ithaca, NY, 14853, USA}}

\author{Huili Grace Xing}
\affiliation{\hbox{Department of Materials Science and Engineering, Cornell University, Ithaca, NY, 14853, USA}}
\affiliation{\hbox{School of Electrical and Computer Engineering, Cornell University, Ithaca, NY, 14853, USA}}
\affiliation{\hbox{Kavli Institute at Cornell for Nanoscale Science, Cornell University, Ithaca, NY, 14853, USA}}

\author{Debdeep Jena}
\affiliation{\hbox{Department of Materials Science and Engineering, Cornell University, Ithaca, NY, 14853, USA}}
\affiliation{\hbox{School of Electrical and Computer Engineering, Cornell University, Ithaca, NY, 14853, USA}}
\affiliation{\hbox{Kavli Institute at Cornell for Nanoscale Science, Cornell University, Ithaca, NY, 14853, USA}}

\maketitle

\textbf{SUPPLEMENTARY MATERIAL}

\setcounter{figure}{0}
\renewcommand{\figurename}{Fig.}
\renewcommand{\thefigure}{S\arabic{figure}}

\vspace{5mm}
\begin{figure*}[h]
\centering\includegraphics[width=1\textwidth]{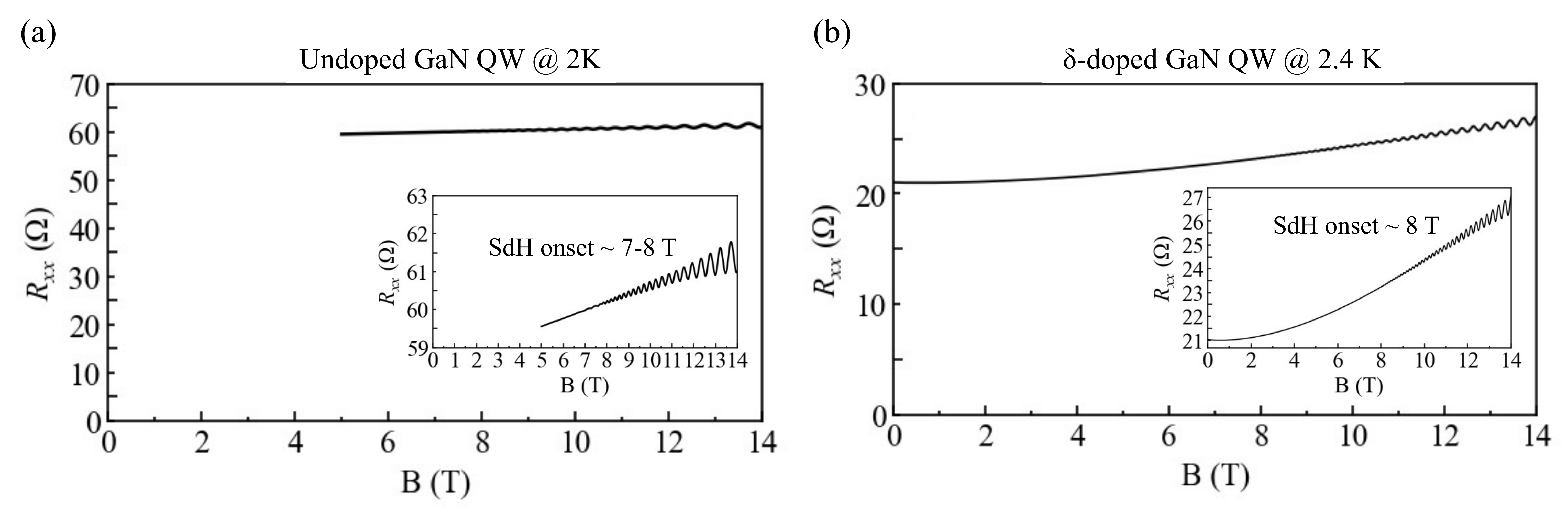}
\caption{\label{figS1} (a) Measured longitudinal magnetoresistance \(R_{xx}\) of 2DEG in the undoped GaN QW from 5 to 14 T at 2 K. Inset shows SdH oscillation onset at around 7-8 T. (b) Measured \(R_{xx}\) of 2DEG in the $\delta$-doped GaN QW from 0 to 14 T at 2.4 K. Inset shows SdH oscillation onset at around 8 T.}
\end{figure*}

\begin{figure*}[h]
\centering\includegraphics[width=0.5\textwidth]{Figure S2.pdf}
\caption{\label{figS2} Measured \(R_{xx}\) of 2DEG in the undoped GaN QW from 9 to 14 T. The green and blue curves represent fourth-order polynomial fits to the SdH oscillation peaks and valleys, respectively. The background resistance (red line) is obtained by averaging the green and blue curves. The oscillatory component \(\Delta R_{xx}\) in Fig. 2(a) is calculated by subtracting the background resistance from the measured \(R_{xx}\). Same method was applied to $\delta$-doped GaN QW.}
\end{figure*}

\begin{figure*}[ht]
\centering\includegraphics[width=1\textwidth]{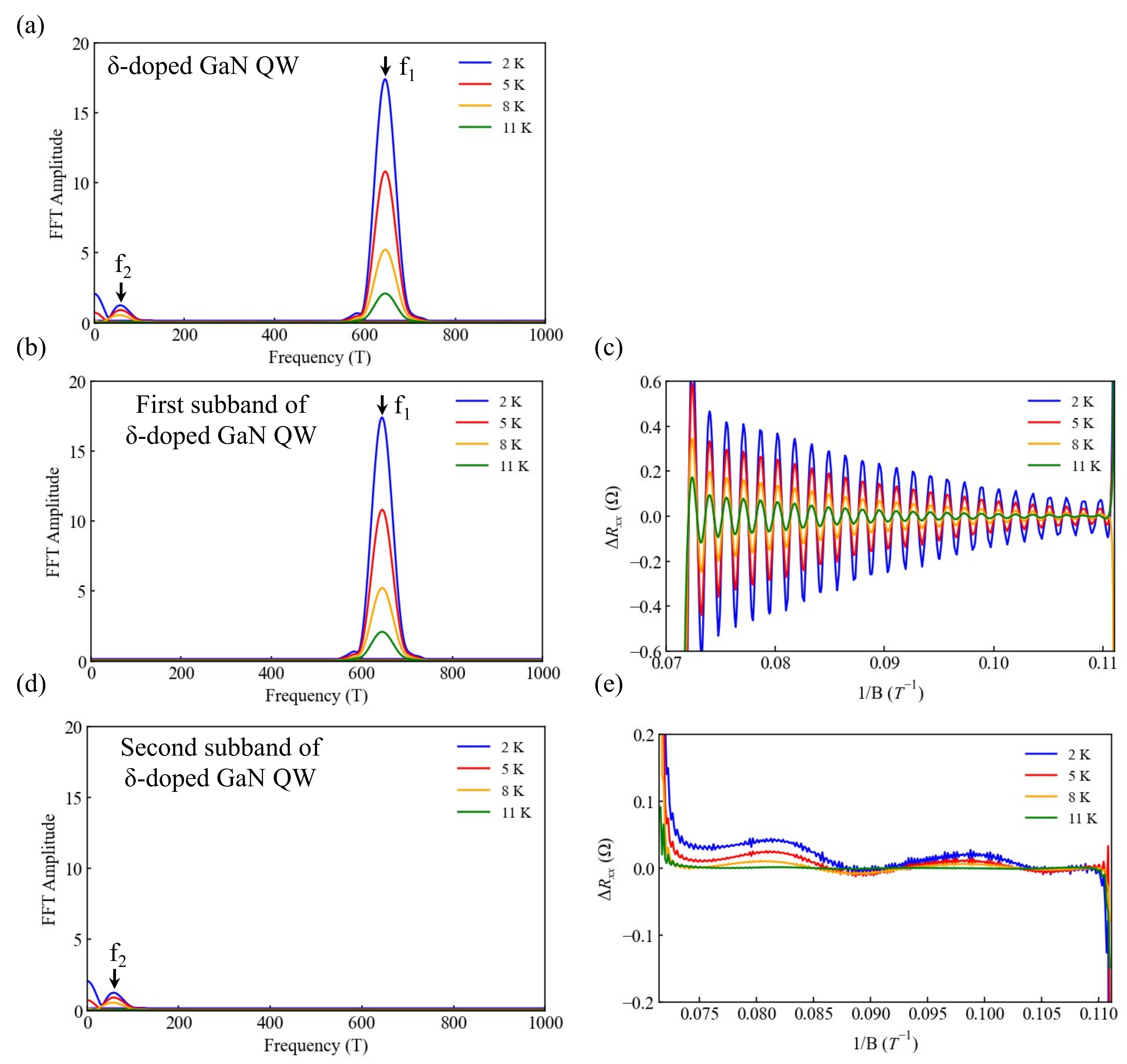}
\caption{\label{figS3} 
(a) FFT spectrum of the $\delta$-doped GaN QW reveals two distinct frequency peaks ($f_1$ and $f_2$), indicating two subband occupation. $f_1$ corresponds to the first subband and $f_2$ to the second subband. Hanning windowing was applied to minimize spectra leakage. 
(b) FFT spectrum of the first subband ($f_1$) of $\delta$-doped GaN QW, where inverse fast Fourier transform (IFFT) was applied to isolate the oscillatory components \(\Delta R_{xx}\) associated with this subband.
(c) The oscillatory component \(\Delta R_{xx}\) plotted as a function of \(1/B\) for the first subband of $\delta$-doped GaN QW. No clear beating pattern is discernible. The peak values of \(\Delta R_{xx}\) were used to extract the electron effective mass and quantum lifetime. The edge artifacts in \(\Delta R_{xx}\) are likely due to the windowing function and edge discontinuities. 
(d) FFT spectrum of the second subband ($f_2$), where a similar analysis was performed.
(e) Oscillatory component \(\Delta R_{xx}\) plotted against \(1/B\) for the second subband of $\delta$-doped GaN QW. }
\end{figure*}

%\section*{References}
%\bibliography{reference}% Produces the bibliography via BibTeX.